\documentclass[journal]{IEEEtran}

\usepackage{graphicx}
\usepackage{subcaption}
\usepackage{balance}
\usepackage{comment}
\usepackage{amsmath}
\usepackage{algorithmic}
\usepackage[ruled,linesnumbered]{algorithm2e}
\usepackage[bookmarks=false]{hyperref}

\ifCLASSINFOpdf
\else
\fi

\usepackage{balance}

\usepackage{multirow}
\usepackage[table]{xcolor}
\begin{document}
	\title{Energy-Efficient Runtime Adaptable L1 STT-RAM Cache Design}


	\author{Kyle Kuan,~\IEEEmembership{Student Member,~IEEE} and Tosiron~Adegbija,~\IEEEmembership{Member,~IEEE}

\thanks{The authors are with the Department of Electrical and Computer Engineering, The University of Arizona, USA, e-mail: \{ckkuan, tosiron\}@email.arizona.edu.}
}
\maketitle

\begin{abstract}

Much research has shown that applications have variable runtime cache requirements. In the context of the increasingly popular Spin-Transfer Torque RAM (STT-RAM) cache, the retention time, which defines how long the cache can retain a cache block in the absence of power, is one of the most important cache requirements that may vary for different applications. In this paper, we propose a \textit{Logically Adaptable Retention Time STT-RAM (LARS)} cache that allows the retention time to be dynamically adapted to applications' runtime requirements. LARS cache comprises of multiple STT-RAM units with different retention times, with only one unit being used at a given time. LARS dynamically determines which STT-RAM unit to use during runtime, based on executing applications' needs. As an integral part of LARS, we also explore different algorithms to dynamically determine the best retention time based on different cache design tradeoffs. Our experiments show that by adapting the retention time to different applications' requirements, LARS cache can reduce the average cache energy by 25.31\%, compared to prior work, with minimal overheads.   

\end{abstract}



\section{Introduction}

The memory hierarchy remains one of the most important components of computer systems, including mobile devices, embedded systems, desktop computers, servers, etc. On-chip caches are especially important for bridging the persistent processor-memory performance gap in computers. The cache subsystem can consume up to 50\% of a processor's total power \cite{mittal14}. As a result, much research has focused on optimization techniques to reduce cache energy consumption without degrading performance. 

An emerging and increasingly popular optimization involves using the non-volatile Spin-Transfer Torque RAM (STT-RAM) instead of traditional SRAM caches. STT-RAM offers several advantages, including non-volatility, higher storage density than SRAM, low leakage power, and compatibility with CMOS technology \cite{Jog12,Dong08,Sun11}. Much prior research and various prototypes, including a few commercial offerings, demonstrate the growing interest in STT-RAMs and their benefits \cite{apalkov13,chung16,park12}. However, dynamic operations in STT-RAM caches accrue significant overheads, compared to SRAM caches, due to long write latency and high dynamic write energy \cite{Sun11}. Furthermore, in resource-constrained general purpose systems (e.g., smartphones, tablets) that execute a variety of applications, which may be unknown a priori, cache requirements typically vary during runtime. As such, a single design-time configuration may be over- or under-provisioned for the runtime execution needs of different applications, thus limiting energy optimization potential. Targeting the L1 cache, due to its high number of dynamic operations, this paper aims to mitigate the energy overheads of STT-RAM caches. We also aim to realize STT-RAM caches that can satisfy variable runtime application needs without introducing substantial overheads. 

To address the aforementioned challenges, we took a close look at STT-RAM's data retention time---the duration for which data is retained in the absence of an external power source. STT-RAM was originally developed to retain data for up to ten years in the absence of an external power source \cite{Diao07}. However, prior work \cite{CongXu11} has revealed that such long retention times also mean that substantially more latency and energy is consumed during writes. These additional overheads could be prohibitive in resource-constrained computer systems like mobile devices. Furthermore, such long retention times are usually unnecessary since most applications' data blocks only need to remain in the cache for no more than one second \cite{Jog12}. Therefore, to reduce the write latency and energy, the retention time can be substantially relaxed, such that it is just sufficient to hold cached data. 


Much prior work has explored the benefits of substantially relaxing the retention time \cite{Smullen11,Sun11,Jog12,Li13,Rodriguez14}. In order to reap the full energy and latency benefits of relaxing the retention time, the STT-RAM cache is sometimes designed such that the relaxed retention time is shorter than several data blocks' lifetimes---i.e., the duration for which the data blocks must remain in the cache. As such, premature eviction of data blocks must be prevented using techniques such as the \textit{dynamic refresh scheme (DRS)} \cite{Smullen11,Sun11,Jog12}. DRS is a DRAM-style mechanism that monitors data blocks' lifetimes and continuously refreshes the blocks that must remain in the cache beyond the retention time. However, the refreshes can incur substantial overheads resulting from the multiple read and write operations required for each refresh operation \cite{Li13}. As such, the optimization potential of the STT-RAM cache may be limited by such schemes.

To mitigate the DRS overheads, a few techniques have been proposed to reduce the number of refresh operations \cite{Li13,Qiu16}. The key drawback of these techniques, however, is that they typically rely on compiler-based data rearrangement. These compiler-based techniques incur overheads due to the increased compilation time and the costs of extra physical circuits to implement the techniques \cite{Li13,Rodriguez14}. In addition, our analysis shows that different applications may require different retention times based on the applications' execution characteristics (e.g., cache block lifetimes). Prior techniques typically feature a single retention time throughout the cache's lifetime \cite{Li13,Qiu16}, and therefore, cannot be dynamically adapted to different applications' runtime needs.

We have extensively analyzed the behavior of cache blocks and their sensitivity to retention times. Based on these analysis, we made three major observations that motivate the work proposed herein. First, similar to other cache configurations (size, line size, associativity), different applications may require different retention times for energy-efficient execution, depending on the applications' cache block characteristics. Second, even though a shorter retention time consumes less energy than a longer one, the longer retention time may be more energy-efficient in the light of the refresh overheads incurred when using shorter retention times. Third, and conversely to the second observation, we also observed that the shorter retention time may be beneficial for some applications, if the reduced retention time does not excessively increase the cache misses. 

Based on our observations, we propose that a relaxed retention time STT-RAM cache's access energy can be substantially reduced by dynamically adapting the retention time to different applications' runtime needs. However, the retention time is an inherent physical characteristic of STT-RAMs \cite{Diao07} that cannot be easily changed during runtime (unlike other cache configurations, such as cache size or associativity \cite{zhang03}). Therefore, we explore our idea of exploiting STT-RAM's density characteristics to logically adapt the retention time to different applications' runtime needs. 

In this paper, we propose \textbf{\textit{Logically Adaptable Retention time STT-RAM (LARS)}} as a practical approach for enabling STT-RAM caches whose retention times can be dynamically adapted to different application requirements. LARS leverages STT-RAM's high density and small area compared to SRAM. A LARS cache comprises of multiple STT-RAM cache units, with only one unit being used at any given time based on the executing application's retention time needs. 

Our major contributions are summarized as follows:  

\begin{description} 
\item[$\bullet$] We propose dynamically adaptable retention time as a practical approach to energy-efficient STT-RAM caches that can satisfy variable runtime application requirements. To this end, we explore our idea of logical adaptation and its potentials for reducing energy, with minimal overheads.
\item[$\bullet$]We analyze both instruction and data cache behaviors for different retention times, and for different applications. Based on our analysis, we show that adaptable retention time provides substantial energy benefits for the data cache, whereas, a static [carefully chosen] retention time suffices for the instruction cache. 
\item[$\bullet$] Based on our analysis of cache behaviors, we explore and evaluate simple and easy-to-implement algorithms to dynamically determine the best retention times during runtime.
\item[$\bullet$]We compare LARS to both SRAM and prior related work (using the widely adopted DRS) to investigate the potentials of LARS and any concomitant overheads. Our experiments reveal that, compared to DRS, LARS reduced the average STT-RAM cache energy by 25.31\%, with negligible increase in the latency and minimal area overheads. 
\end{description}

\vspace{-3pt}
\section{Background and Related Work}

\begin{figure}[b]
		\vspace{-15pt}
		\centering
		\includegraphics[width=.55\linewidth]{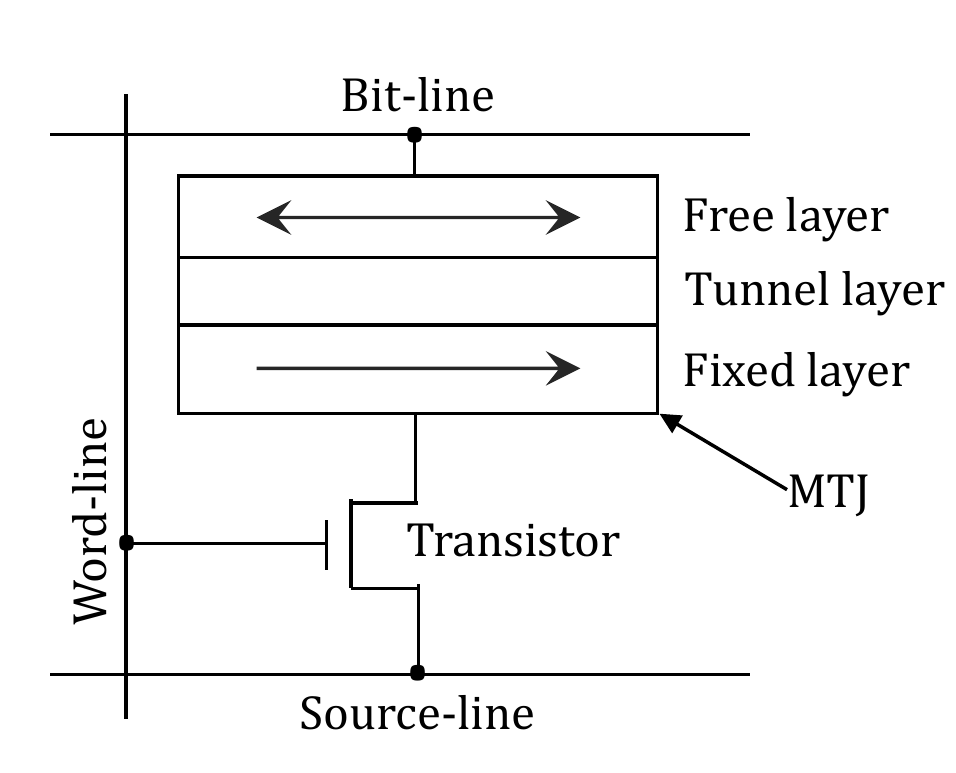}
		\vspace{-7pt}
		\caption{STT-RAM basic cell structure}
		\vspace{-7pt}
		\label{fig:sttram}
	\end{figure}

Fig. \ref{fig:sttram} illustrates the STT-RAM's basic cell structure. STT-RAM uses a magnetic tunnel junction (MTJ), which contains two ferromagnetic layers separated by an oxide barrier/tunnel layer, as the binary storage cell. Similar to other resistive memories, STT-RAM uses non-volatile, resistive information storage in a cell. The MTJ's ferromagnetic layers are a free layer and a fixed layer---wherein the free layer's direction with respect to the fixed layer (parallel or anti-parallel) indicates the cell's "0"/"1" state. Details of the STT-RAM's structure are presented in \cite{Chun13}. In this section, we present a brief overview of related prior work on volatile STT-RAMs that provides the background for LARS.

\subsection{Refresh Schemes on Volatile STT-RAM Cache}
Prior work has shown that \textit{volatile STT-RAMs} featuring a relaxed/reduced retention time can significantly reduce the write energy and latency \cite{Smullen11,Sun11,Jog12}. The retention time can be relaxed by reducing the MTJ's planar area \cite{Smullen11,Jog12} or by reducing the MTJ's saturation magnetization \cite{Sun11}. To achieve maximum benefits, this reduction in retention time must be substantial (e.g., from 10 years to a few seconds). As such, in several cases, cache blocks may still be referenced beyond the retention time. To prevent data loss in volatile STT-RAMs, Sun et al. \cite{Sun11} proposed the \textit{dynamic refresh scheme} (DRS), which uses DRAM-style refreshes to maintain data correctness for blocks that must remain in the cache beyond the retention time. DRS features a counter to monitor each block's lifetime in relation to the retention time. When the retention time elapses, the cache controller continuously refreshes the cache block until its lifetime expires (e.g., through eviction). DRS has been used in more recent work in different forms/implementations, and without loss of generality, we collectively call these techniques DRS.

DRS incurs energy overheads due to the refresh operations, which could be significantly large in some applications. To reduce the refresh overheads, Jog et al. \cite{Jog12} proposed the \textit{cache revive} scheme, a flavor of DRS. In cache revive, a small buffer is used to temporarily store cache blocks that prematurely expired due to elapsed retention time. Most recently used (MRU) blocks are then copied back into the cache and refreshed. 

To further reduce the refresh overheads, more recent works used compiler-based techniques---such as code optimization \cite{Li13} and loop scheduling \cite{Qiu16}. Some techniques attempt to reduce the write energy by reducing the number of unnecessary writes. For example, Bouziane et al. \cite{Bouziane18} leveraged the compiler to identify redundant writes, known as "silent-stores." These redundant writes were then prevented from occurring in order to reduce the write energy. However, these works preclude runtime optimization and incur overheads, since they feature a single retention time and rely on dedicated hardware to deal with the data loss in volatile STT-RAM cache.  

	\begin{figure}[b]
		\vspace{-15pt}
		\centering
		\includegraphics[width=.6\linewidth]{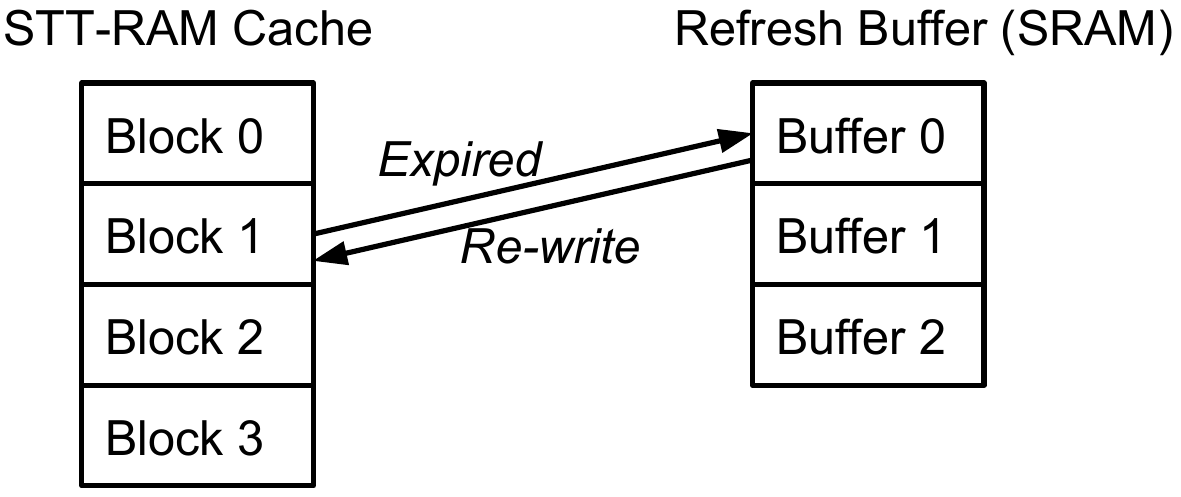}
		\caption{Overheads of dynamic refresh scheme}
		\vspace{-7pt}
		\label{fig:refresh}
	\end{figure}
    
\subsection{Cost of Refresh Schemes and Motivation for LARS}

Fig. \ref{fig:refresh} illustrates some of the overheads incurred by the dynamic refresh scheme using three cache blocks. Assuming that the STT-RAM cache's retention time has elapsed, but Block 1's lifetime has not, DRS refreshes the block by copying it into a refresh buffer---the buffer can be SRAM or STT-RAM---and then writing the block back into the STT-RAM cache. Each refresh operation to transfer the block between the STT-RAM cache and the buffer costs: 1) an STT-RAM read; 2) a buffer write; 3) a buffer read and; 4) an STT-RAM write. Considering that most applications feature several refreshes throughout execution, the energy overheads of these operations can be prohibitive \cite{Li13}, especially in resource-constrained systems. 

	\begin{figure}[t]
		\centering
		\includegraphics[width=.7\linewidth]{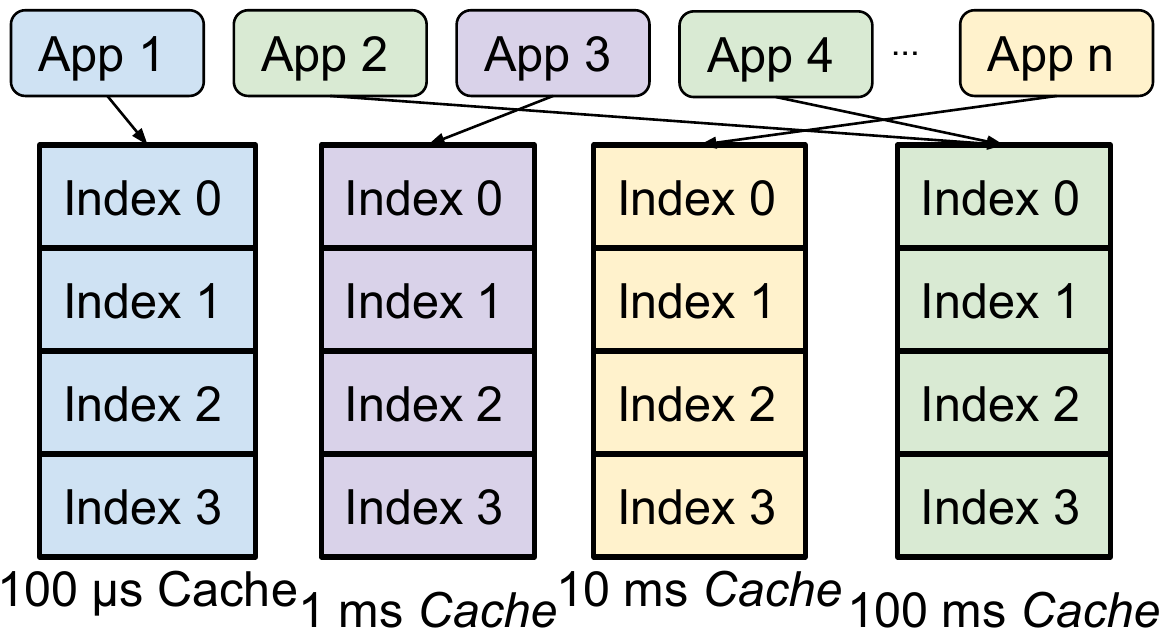}
		\caption{Adapting STT-RAM retention times to applications' requirements}
		\vspace{-10pt}
		\label{fig:adapt}
	\end{figure}

Jog et al. \cite{Jog12} studied the cache block lifetimes of different applications, with respect to the last level cache (LLC), to reveal that a retention time of 10$ms$ with a buffer sufficed for the range of applications considered. Our analysis further revealed that, based on the variable block lifetimes in different applications, retention times can be adapted to the applications to reduce the access energy and latency. This insight forms the basis of the work presented herein.  

\subsection{Other Improvements in the use of STT-RAM}
While this paper focuses on the L1 cache, prior work has provided insights on STT-RAM's limitation throughout the memory hierarchy. To this end, several techniques have been proposed to address the challenges posed by STT-RAMs, especially with respect to the write energy and latency overheads. Ranjan et al. \cite{Ranjan17} used approximated storage to provide an energy-efficient solution to organize STT-RAM in the level two (L2) cache. Reed et al. \cite{Reed17} used conditional random replacement to mitigate repeated writes on some cache blocks and improve write lifetime of STT-RAM L2 cache. For the last level cache (LLC), the most common works emphasize predicting and balancing the usability of cache blocks in order to avoid the insertion of non-reused cache blocks in the LLC. In effect, the number of writes is reduced, while also reducing energy consumption and improving the cache endurance \cite{RR17} \cite{DASCA} \cite{Benzene} \cite{Korgaonkar18}. Other work on LLC (e.g., Zeng et al. \cite{Zeng17}) quantified data block replacements based on the required write activities, and used minimum bit transitions to replace cache blocks, in order to reduce the write energy. 

\section{Logically Adaptable Retention Time STT-RAM (LARS) Cache}
Fig. \ref{fig:adapt} illustrates the overarching idea of how LARS works. Consider a set of applications running on a general-purpose system (e.g., smartphone), where many of these applications may be unknown at design time. The cache is designed such that it has a set of retention times that can satisfy different applications' runtime requirements. (Section \ref{sec:miss} details how these requirements can be determined). As illustrated in Fig. \ref{fig:adapt}, based on their cache block characteristics, $App1$ requires a 100 $\mu$s retention time, $App2$ and $App4$ require a 100 ms retention time, and so on. To specialize the retention times to the application requirements, in order to reduce energy consumption, LARS allows each application to execute on a cache unit with a retention time that best satisfies the application's needs, given the constrained design space of retention times. To facilitate this runtime adaptability, LARS also involves a hardware structure that dynamically determines the best retention time for each executing application. This section motivates LARS through an analysis of applications' retention time behaviors, and details LARS architecture, algorithms, and overheads. 

\subsection{Retention Time Analysis}\label{sec:miss}

To motivate our work, we analyzed how retention times affect applications' cache miss rates. We used cache miss rates as an indicator of the cache's performance for the executing application. Ideally, for energy efficiency, we would want the smallest retention time that satisfies an application's cache requirements without substantially trading off the performance. 

\begin{figure}[t]
\centering
\begin{subfigure}[t]{.49\textwidth}
  \centering
  \includegraphics[width=0.8\linewidth]{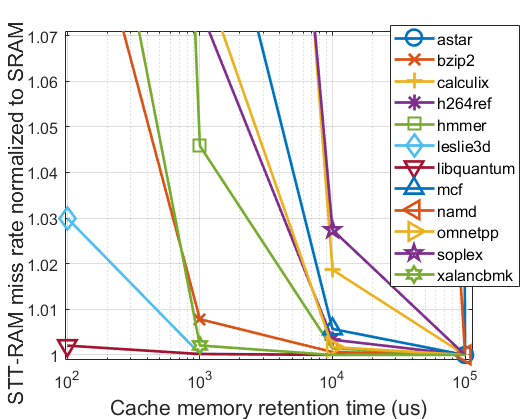}
  \caption{Data cache}
  \label{fig:dcache_miss}
\end{subfigure}%

\begin{subfigure}[t]{.46\textwidth}
  \centering
  \includegraphics[width=0.83\linewidth]{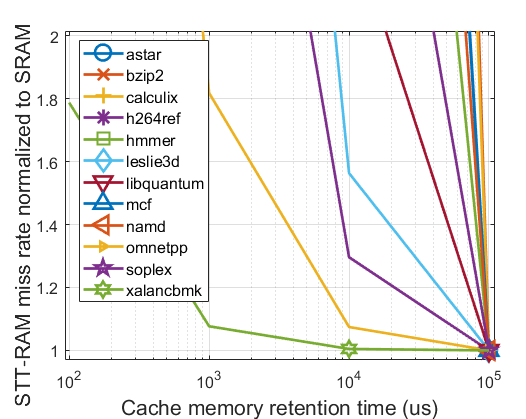}
  \caption{Instruction cache}
  \label{fig:icache_miss}
\end{subfigure}
\caption{STT-RAM cache miss rate changes for different retention times (normalized to SRAM, baseline of 1)}
\label{fig:miss}
\end{figure}

Fig. \ref{fig:miss} illustrates the relationship between cache miss rates and retention times for different applications in the SPEC 2006 \cite{spec2006} benchmark suite. Our experimental setup is detailed in Section \ref{sec:setup}. The miss rates for the different STT-RAM retention times are normalized to the applications' SRAM miss rates with the same base cache configuration (32KB, 4-way set associative, 64B line size). Since a higher retention time implies higher energy and latency, our goal was to explore the lowest retention times that maintained comparable cache miss rates to the SRAM (baseline of one in the figure). We determined the best retention time as one that achieved miss rates within a threshold of 5\% compared to SRAM. We empirically determined this threshold by observing that this change in miss rates did not result in observable change in energy consumption or performance.

In general, as expected, the miss rates decreased as the retention times increased for all the applications. However, for the different applications, there were variances in the benefit of further increasing the retention time beyond certain amounts. Furthermore, we also observed that the data and instruction caches behaved differently with respect to the retention time. For the data cache, the retention times that achieved low cache miss rates varied for the different benchmarks. As shown in Fig. \ref{fig:dcache_miss}, $libquatum$'s and $leslie3d$'s best retention time were 100$\mu$s. $hmmer$'s, $bzip2$'s and $xalancbmk$ were 1ms. $astar$'s and $namd$'s were 100ms. For the other five benchmarks, the best retention time was 10ms, which we also found to be an average good retention time across all the benchmarks (similar to prior work \cite{Jog12}). However, our observations revealed that using the single averagely good retention time of 10ms limited the optimization potential for each individual application, and in effect, for all the applications on average. Compared to using the best retention time for each application, using a static 10ms retention time increased the average data cache miss rates by 18.45\%. 

\begin{figure}[b]
\centering
\includegraphics[width=3.8in]{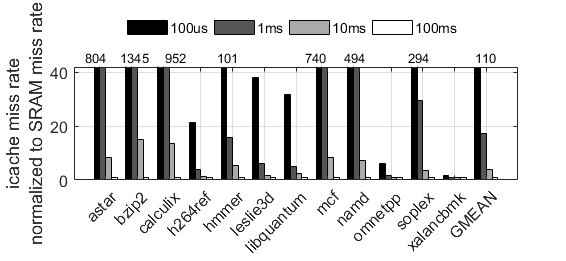}
\vspace{-7pt}
\caption{Instruction cache miss rates for the different retention times normalized to SRAM}
\label{fig:icache_miss_compare}
\end{figure}

In general, instructions typically exhibit less runtime variability than data. As such, for the instruction cache, we observed much less variability in the retention time requirements of different benchmarks. As depicted in Fig. \ref{fig:icache_miss}, the 100ms retention time was best for eleven of the twelve benchmarks considered. Only $xalancbmk$ required a different retention time of 10ms. We also observed that using a smaller retention time than 100ms resulted in substantial increases in the miss rates for most of the benchmarks. Fig. \ref{fig:icache_miss_compare} illustrates this observation. For instance, reducing the retention time to 10ms increased the miss rates by more than 2x for eight out of twelve benchmarks. Reducing the retention time to 100$\mu$s astronomically increased the miss rates by 1345x and 952x for $bzip2$ and $calculix$, respectively. Therefore, we concluded that unlike the data cache, adaptable retention time would bear no benefits for the instruction cache. We decided to use a static 100ms retention time for the instruction cache in the rest of our experiments.   

\subsection{LARS Architecture} \label{sec:implementation}
Prior work has shown that STT-RAM is 3 to 9 times denser than SRAM \cite{Sun11,Jog12}. However, due to design constraints, such as access latency considerations, L1 cache sizes are usually limited (e.g., 16 -- 32KB in modern-day smartphones). Thus, thanks to its density, an STT-RAM would take up a much smaller physical area than SRAM for the same memory capacity. Leveraging STT-RAM's physical density advantages, we propose a LARS cache that comprises of four STT-RAM units, which will take up approximately the same physical area as one SRAM cache of the same memory capacity. Even though we have empirically determined that these four units suffice for our benchmarks, designers can opt to use a different number of STT-RAM units depending on the target applications or application domains. 

We note that prior work has shown that MTJ cells can have reliability issues (e.g., read disturbance) \cite{Kang15,Kang17}, particularly in relaxed retention time MTJs \cite{Smullen11}. While there is much ongoing work to tackle these issues, they are beyond the scope of this paper. For our work, we assume that the STT-RAM caches can be fabricated with the desired retention times. Each STT-RAM cache features different STT-RAM 'units' implemented as functionally self-contained physical banks within the STT-RAM chip. Details of how we modeled and simulated this cache are in Section \ref{sec:setup}.

\begin{figure}[t]
	\centering
	\includegraphics[width=.9\linewidth]{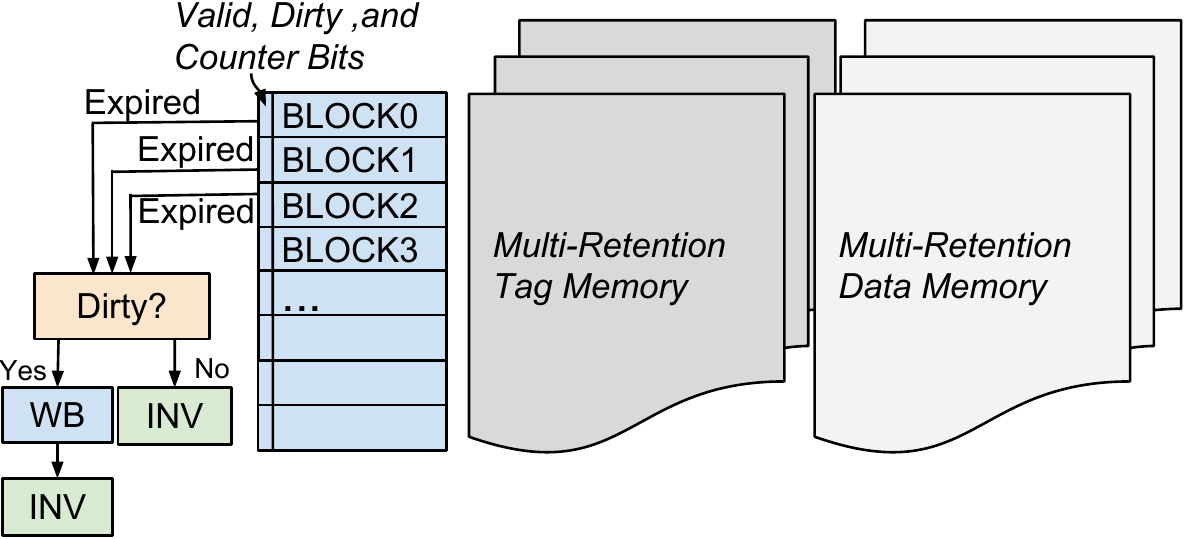}
	\vspace{-7pt}
	\caption{LARS architecture}
	\vspace{-10pt}
	\label{fig:lars}
	\end{figure}

Fig. \ref{fig:lars} depicts the proposed LARS architecture, which comprises of four STT-RAM units with four different data memory retention times. As previously alluded to in Section \ref{sec:miss}, these retention times are empirically determined at design time to satisfy a range of application needs, depending on the target system. The cache also comprises of four tag memory units, a per-block status array wherein each element contains a valid bit, dirty bit (assuming a write-back cache) and \textit{monitor counter} bits. For each application, the cache controller only accesses a single cache unit at a given time, depending on the application's retention time needs. Even though LARS eliminates the need to refresh cache blocks, the monitor counter determines when to eliminate an expiring cache block (e.g., through invalidation) in order to prevent data corruption resulting from a prematurely elapsed retention time. We designed the monitor counter similarly to \cite{Sun11}, and assume the counter's clock period is $N$ times smaller than the retention time. That is, when a block's counter reaches $N-1$ (starting from 0), the block has reached the maximum retention time and should be invalidated. Before eliminating the block, its dirty bit must be checked to determine whether or not it must be written back to main memory, as in a normal write-back cache.

\begin{figure}[b]
 	\vspace{-15pt}
    \centering
	\includegraphics[width=3in]{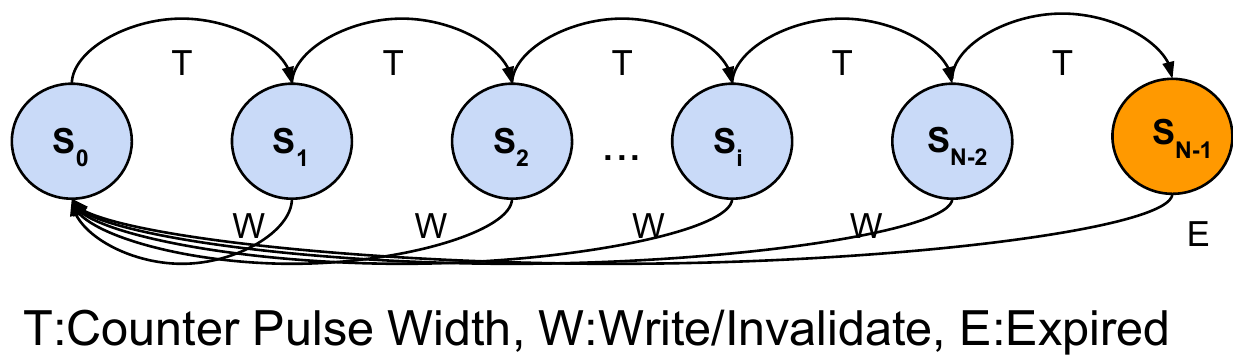}
	\vspace{-7pt}
	\caption{Monitor counter state machine for each cache block}
	\vspace{-7pt}
	\label{fig:counter}
    \vspace{-4pt}
	\end{figure}

Fig. \ref{fig:counter} shows the N-bit monitor counter's state machine. The state machine comprises of states $S_0$ to $S_{N-1}$, which advance on the monitor clock's rising edge. Within each state, the counter resets to $S_0$ if the cache block receives a write or invalidate request. At state $S_{N-1}$, the counter sets the $E$ signal, which triggers LARS (via the cache controller) to check the block's dirty bit. If the block is dirty, the block is written to the main memory. Otherwise, the block is invalidated. Note that LARS requires minimal modifications to the cache controller, since these processes (e.g., writing back/invalidating a cache block) are implemented in state-of-the-art cache controllers. Our analysis shows that the counter comprises little overhead (details in Section \ref{overheads}).
    
\subsection{Determining the Best Retention Time} \label{sec:algorithm}

We assume that the cache controller orchestrates the 'powering on/off' of the appropriate STT-RAM units. Since STT-RAMs work on the principle of \textit{normally-off computing} \cite{ando14}, the cache controller orchestrates the process by simply writing/reading applications' data blocks to/from the appropriate STT-RAM units. Thus, LARS does not require any modifications to the cache controller beyond a 2-bit \textit{location array} to indicate which STT-RAM unit to use for an executing application. Using more or fewer STT-RAM units would change the number of bits required for the location array.

To non-intrusively determine the best retention time for different applications---and in effect, which STT-RAM unit to use---we designed a low-overhead hardware \textit{LARS tuner} to implement the algorithms described herein (the tuner overheads are described in Section \ref{overheads}). To minimize the runtime tuning overhead, we explored different techniques for dynamically determining an application's best retention time. To enable easy/low-overhead implementation, we chose to use simple algorithms. We found, during our experiments, that these simple algorithms sufficed for achieving LARS' full benefits. We considered three approaches: a sampling technique, which samples all the retention time units to determine the best retention time, and two tuning algorithms, which we call \textit{LARS-Optimal} and \textit{LARS-Miss}. Both algorithms achieve different tradeoffs with respect to tuning accuracy and implementation overheads. In what follows, we describe these different approaches.

\subsubsection{Sampling Technique} \label{sec:sampling}

We first explored a simple sampling technique that exhaustively samples every available retention time to determine the \textit{best} retention time. The application is sampled on each STT-RAM unit for a \textit{tuning interval} during which the energy consumption is measured. We used intervals of 100million instructions, which we determined to be sufficient time to gather stable statistics of the application's execution. After all the retention times have been sampled, the best (e.g., lowest energy) retention time is then selected and stored in a low-overhead data structure for subsequent use. 

Note that given the constrained design space of four retention times, the tuning overheads with respect to time and energy are not prohibitive. Since we used a tuning interval of 100 million instructions, tuning only takes a small fraction of each application's execution, after which the best retention time is used for the rest of the execution. The tuning overheads are rapidly amortized over the application's execution, which can span trillions of instructions \cite{spec2006}. Alternatively, time intervals (e.g., in seconds or cycles) can be used for the tuning intervals. Shorter tuning intervals can also be used for more fine-grained tuning (e.g., at a phase granularity wherein retention times are specialized for different phases within an application). We leave exploration of phase-based LARS for future work. 

We also explored different objective functions (energy, latency, or energy-delay product (EDP)) for evaluating the best retention time during runtime. We found that using the EDP provided a Pareto-optimal balance between energy and latency optimization, as compared to using latency or energy as the objective function (details in Section \ref{sec:samplingResult}). Thus, we used EDP as the objective function in describing the LARS algorithms and in our experiments.   

\subsubsection{LARS-Optimal} \label{subsec:LARS-Optimal}
For easy practical implementation, we designed LARS-Optimal as a simple heuristic/algorithm to determine the best retention times during runtime. The algorithm determines the best retention time using a cache energy model \cite{Tosi15} based on the number of cache accesses, writebacks, misses, and the associated latencies.

\begin{algorithm}[t]
\SetDataSty{small}
\SetKwData{BaseEDP}{BaseEDP}
\SetKwData{CurEDP}{CurEDP}
\SetKwData{OutputRetentionTime}{OutputRetentionTime}
\SetKwFunction{samplingEDP}{samplingEDP}
\KwData{Retention time set  $R=\{100 \mu s,1ms,10ms,100ms\}$}
\KwResult{Best retention time}
\BlankLine
\emph{\BaseEDP $\leftarrow$ \samplingEDP{$100ms$}}\;
\emph{\OutputRetentionTime $\leftarrow 100ms$}\;

\ForEach{$r \in R-\{100ms\}$}{
  \emph{\CurEDP $\leftarrow$ \samplingEDP{$r$}}\;
  \If{\CurEDP $=<$ \BaseEDP}{ \label{line:optimalCond1}
    \BaseEDP $\leftarrow$ \CurEDP\; \label{line:updateEDP}
    \OutputRetentionTime $\leftarrow r$\; 
    }
  \Else{
    \Return{\OutputRetentionTime, \BaseEDP}\;
  }\label{line:optimalCond2}
}
\Return{\OutputRetentionTime, \BaseEDP}\;

\caption{LARS-Optimal Tuning Algorithm}
\label{algo:optimal_algo1}
\end{algorithm}

\begin{algorithm}[t]
\SetDataSty{small}
\SetKwData{BaseMisses}{BaseMisses}
\SetKwData{CurMisses}{CurMisses}
\SetKwData{CurMissRate}{CurMissRate}
\SetKwData{OutputRetentionTime}{OutputRetentionTime}
\SetKwFunction{samplingMisses}{samplingMisses}
\KwData{Retention time set  $R=\{100 \mu s,1ms,10ms,100ms\}$}
\KwResult{Best retention time}
\BlankLine
\emph{\BaseMisses $\leftarrow$ \samplingMisses{$100ms$}}\;
\emph{\OutputRetentionTime $\leftarrow 100ms$}\;

\ForEach{$r \in R-\{100ms\}$}{
    \emph{\CurMisses,\CurMissRate $\leftarrow$ \samplingMisses{$r$}}\;
    \If{LARS-Miss-LB is true \&\& \CurMissRate $<$ 0.05\%}{\label{line:LBCond1}
        \OutputRetentionTime $\leftarrow r$\;    
    }
    \ElseIf{\CurMisses $< \BaseMisses*1.05$}{ \label{line:MissCond1}
        \OutputRetentionTime $\leftarrow r$\;
        
    }
    \Else{
        \Return{\OutputRetentionTime, \BaseMisses}\;
    }\label{line:MissCond2}
}
\Return{\OutputRetentionTime, \BaseMisses}\;

\caption{LARS-Miss/-LB tuning algorithm}
\label{algo:miss_algo1}
\end{algorithm}

\begin{algorithm}
\SetDataSty{small}
\SetKwData{BaseMisses}{BaseMisses}
\SetKwData{CurMisses}{CurMisses}
\SetKwData{BaseEDP}{BaseEDP}
\SetKwData{CurRetentionTime}{CurRetentionTime}
\SetKwData{CurEDP}{CurEDP}
\SetKwData{ReTuneApp}{ReTuneApp}
\SetKwFunction{samplingMisses}{samplingMisses}
\SetKwFunction{samplingEDP}{samplingEDP}
\KwData{\BaseEDP, \BaseMisses, \CurRetentionTime}
\KwResult{\ReTuneApp}
\BlankLine
\emph{\ReTuneApp $\leftarrow false$}\;
\If{LARS-Miss-LB is true or LARS-Miss is true}{
    \CurMisses $\leftarrow$ \samplingMisses{\CurRetentionTime}\;
    \ReTuneApp $\leftarrow (\CurMisses > \BaseMisses*1.05) $\;
    }
\ElseIf{LARS-Optimal is true}{
    \CurEDP $\leftarrow$ \samplingEDP{\CurRetentionTime}\;
    \ReTuneApp $\leftarrow (\CurEDP > \BaseEDP*1.05) $\;
    }
\Return{\ReTuneApp}\;

\caption{Checking process}
\label{algo:checking}
\end{algorithm}

Algo. \ref{algo:optimal_algo1} depicts the LARS-Optimal tuning algorithm, which runs during an application's first execution. The algorithm takes as input the retention time set, and outputs the application's best retention time. When the application begins, LARS defaults to the maximum retention time. The application is then executed for a \textit{tuning interval}, during which the execution statistics are collected from hardware performance counters and the EDP is calculated using the energy model. For our experiments, we used a tuning interval of 100 million instructions to provide a balanced tradeoff between tuning overhead and accuracy; this interval, however, can be adjusted based on specific system tradeoffs \cite{gordon-ross07}.

	\begin{figure}[b]
	    \vspace{-15pt}
		\centering
		\includegraphics[width=\linewidth]{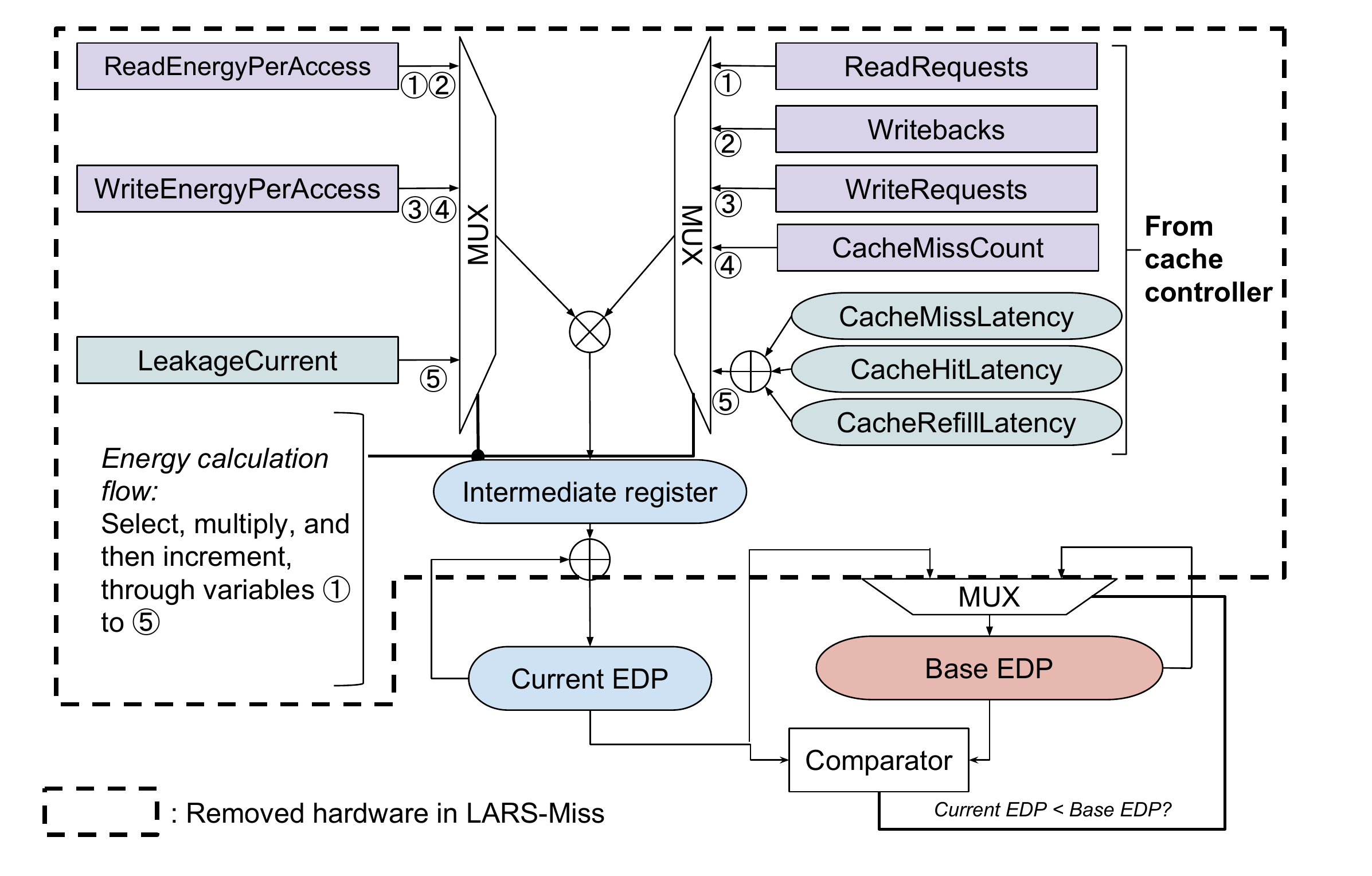}
		\vspace{-7pt}
		\caption{Datapath for the energy model}
		\vspace{-4pt}
		\label{fig:datapath}
	\end{figure}

Fig. \ref{fig:datapath} illustrates our datapath, which implements the energy model for calculating the energy. To calculate the cache access latency, the datapath uses the cache miss latency, hit latency, and refill latency, which can be derived from the number of misses, hits, and total accesses. These statistics can be obtained from the processor's hardware performance counters, which are featured in most state-of-the-art processors. The datapath contains a multiply-accumulate (MAC) unit, comprising of a multiplier, \textit{intermediate register}, and adder, to calculate the current energy and EDP. The circled numbers in Fig. \ref{fig:datapath} represent the order in which the controller state machine selects data items for the MAC. Since LARS uses the EDP as the objective function, as alluded to in Section \ref{sec:sampling}, the calculated \textit{current EDP} is stored as the \textit{base EDP} for comparison during the tuning process. 

As shown in Algo. \ref{algo:optimal_algo1}, LARS-Optimal iterates through the retention times in descending order by running the application for one tuning interval per retention time. After each iteration, LARS-Optimal compares the current EDP value to the base EDP. If the current EDP is less than or equal to the previous EDP, the current retention time is stored and the base EDP is updated to the current EDP. Otherwise, the previous retention time is retained for the application, after which the tuning process exits (lines~\ref{line:optimalCond1}--\ref{line:optimalCond2}). For non-intrusive operation, the retention time and EDP values are stored in a small low-overhead hardware data structure (Section \ref{overheads}).  

To provide a feedback mechanism to LARS-Optimal, we also included a \textit{checking process} shown in Algo. \ref{algo:checking}. The checking process monitors the EDP to detect any deviations from the expected values (based on the initial tuning process). Such a deviation can occur as a result of new data inputs or changes in execution conditions. If the EDP deviates from the stored value by more than 5\%, LARS re-tunes the retention time for the application.

\subsubsection{LARS-Miss} \label{subsec:LARS-Miss}

As illustrated in Fig. \ref{fig:datapath}, LARS-Optimal incurs hardware overheads resulting from the datapath registers and MAC unit required for runtime energy calculations. Thus, LARS-Miss seeks to reduce these overheads. From the analysis in Section \ref{sec:miss}, which shows the sensitivity of cache miss rates to different retention times, we observed that we could predict the retention times using the applications' cache misses. We established LARS-Miss by assuming that the largest retention time (100ms in our case) has the ideal or closest performance to the SRAM. Thus, LARS-Miss, rather than calculating the EDP during iterations as in LARS-Optimal, only monitors changes in cache miss. 

As shown in Algo. \ref{algo:miss_algo1}, instead of recording parameters like access counts, latencies, or writebacks, only the number of cache misses is recorded from the processor's hardware performance counters. Similar to LARS-Optimal, LARS-Miss begins with the maximum retention time, executes the application for one tuning interval, and sets the recorded number of cache misses as the \textit{base} number of cache misses. 

LARS-Miss iterates through the retention times in descending order, and compares the number of cache misses observed for each interval with the base number of cache misses. For each interval, LARS calculates the difference between the base and current number of cache misses. If the number of cache misses observed with the current retention time does not degrade the number of misses in maximum retention time by more than 5\%, the current retention time is stored. We assigned the base misses as the misses achieved by the longest retention time (typically also the lowest misses). As such, unlike LARS-Optimal, LARS-Miss does not need to store current misses, unless it improves over the base. If current misses exceed base misses by more than 5\%, the previous retention time is retained as the application's valid retention time, and the tuning process exits (lines~\ref{line:MissCond1}--\ref{line:MissCond2}). We empirically determined that 5\% was a sufficient tradeoff between the number of cache misses and the improvement from a smaller retention time. This process continues until the best retention time is determined (i.e., when tuning exits). 

\subsubsection{LARS-Miss-LB} \label{subsec:LARS-Miss-LB} While LARS-Miss improved over the dynamic refresh scheme for most benchmarks, we observed that the dynamic refresh scheme outperformed LARS-Miss for two benchmarks: $astar$ and $namd$. We attribute this behavior to their low initial misses in the base retention time. For these applications, LARS-Miss easily exceeded the 5\% threshold of cache misses during tuning, and thus used a higher retention time, resulting in higher energy consumption. Based on this observation, we also created a flavor of LARS-Miss, called \textit{LARS-Miss-LB}, which operates as follows. If the base retention time's total miss rate is extremely small (less than 0.05\%), the extra energy/latency produced by expiration misses could be offset by choosing a smaller retention time unit. As shown in Algo. \ref{algo:miss_algo1}, if the miss rate is below 0.05\%, LARS-Miss-LB chooses the smaller retention time and continues the iteration. Otherwise, the tuning process continues as described in LARS-Miss.

Using the cache miss for predictions leads to fewer calculations, less hardware resource for storage and computations (the statistic registers and MAC unit are no longer required), and allows for easy runtime measurement from hardware performance counters. The dotted line in Fig. \ref{fig:datapath} illustrates the hardware resources that are eliminated by LARS-Miss, since only the miss rate is measured. However, registers, comparators, and muxes are still required to enable comparisons of the cache miss rate to earlier iterations.

Similar to LARS-Optimal, LARS-Miss also uses the checking process (Algo. \ref{algo:checking}). However, unlike LARS-Optimal, the base value for comparison in LARS-Miss remains fixed as the number of misses achieved by the $largest$ retention time for the executing application.

\subsection{LARS Overheads} \label{overheads}

LARS' main overheads result from 1) the \textit{LARS hardware}, 2) \textit{runtime tuning}, and 3) \textit{switching overheads}. We estimated the hardware overheads using Verilog implementations, synthesized with Synopsys Design Compiler \cite{synopsys}, and the tuning and switching overheads using simulations (detailed in Section \ref{sec:setup}).

The hardware overheads include the monitor counters (Section \ref{sec:implementation}) and the LARS tuner. The tuner implements the LARS-Optimal algorithm (Section \ref{sec:algorithm}), energy calculation datapath (Fig. \ref{fig:datapath}), and storage for retention time and energy histories (Section \ref{sec:algorithm}). The monitor counter requires $n = log_2N$ bits, where $N$ is the number of monitor clock periods. For example, for a 100$\mu$s retention time and 10$\mu$s clock period, $N = 10$, and $n = 4$. A 32KB cache with 64B lines has 512 monitor counters for each STT-RAM unit; each monitor counter requires 4 bits. The monitor counters in our four-unit LARS design constitutes an area overhead of 3.12\% for a 32KB cache. 

We synthesized the LARS-Optimal tuner with SAED\_EDK90 Synopsys standard cell library. The estimated area overhead was 0.0145 $mm^2$, dynamic power was 28.04 mW, and leakage power was 422.68 $\mu$W. With respect to the ARM Cortex-A15 \cite{CortexA15area}, for example, the tuner's overhead is negligible (around 0.095\%).

Both LARS-Optimal and LARS-Miss/LARS-Miss-LB can further reduce tuning overheads since the algorithms do not exhaustively search the retention time design space. In our experiments, LARS-Miss reduced the search time by 18.75\% compared to sampling, while LARS-Optimal and LARS-Miss-LB reduced by 6.25\% and 10.41\%, respectively.

The \textit{switching overhead} is the energy and latency incurred while migrating the cache state (tag and data) from one STT-RAM unit to another during tuning. Switching occurs every time an application is initially executed or when runtime changes to the application's characteristics necessitate a re-tuning. We estimated that in the worst case (for the 100ms retention time), each migration took approximately 4608 cycles and 57.34nJ energy. In the sampling technique (the worst case scenario), the total switching through all the STT-RAM units for each application incurred time and energy overheads of 15872 cycles and 197.12nJ, respectively. In the context of full application execution, the worst case switching energy and latency overheads were infinitesimal and rapidly amortized during execution.

\begin{table*}[ht]
\vspace{-10pt}
\renewcommand{\arraystretch}{1.3}
\caption{Cache parameters of SRAM and STT-RAM with different retention times}
\label{tab:retention}
\centering
\begin{tabular}{c||c|cccc}
    \hline
    Cache Configuration				&\multicolumn{5}{c}{32KB, 64B line size, 4-way}\\
    \hline
    \hline
    Memory Device				&SRAM	    &STT-RAM-100$\mu$s	&STT-RAM-1ms	&STT-RAM-10ms	&STT-RAM-100ms\\
    \hline
    Write Energy (per access) 	&0.033nJ    &0.040nJ	        &0.056nJ        &0.076nJ        &0.101nJ\\
    \hline
    Read Energy (per access)	&0.033nJ    &0.012nJ	        &0.012nJ        &0.011nJ        &0.011nJ\\
    \hline
    Leakage Power               &38.021mW     &\multicolumn{4}{c}{1.753mW}		\\
    \hline
    Hit Latency (cycles)    	&3	        &2 			        &2		        &2 		        &2\\
    \hline
    Write Latency (cycles)    	&3	        &3			        &4		        &5		        &7\\
    \hline
\end{tabular}
\vspace{-15pt}
\end{table*}

\section{Experimental Setup} \label{sec:setup}
To evaluate and quantify the benefits of LARS, we modified the GEM5 simulator \cite{gem5} to model LARS\footnote{The modified GEM5 version can be found at \url{www.ece.arizona.edu/tosiron/downloads.php}}. We added a new retention parameter to GEM5 in order to model the variable retention time behavior. We compared the time at which a block was inserted with the time at which the block was accessed by CPU. We used the modified simulator to obtain the block's lifetime to determine whether the lifetime has exceeded the retention time. If the lifetime exceeds the retention time, the block is invalidated or written back to lower level memory (if dirty). We also implemented DRS \cite{Sun11,Jog12}---the most related work to ours---in GEM5 to enable comparison with prior work. We modeled DRS as a "perfect" refresh scenario, meaning that there were no extra misses caused by failed refreshes, there were no unnecessary refreshes, and there were no refresh-related latency overheads. We modeled DRS as such to reflect the best case scenario for DRS in order to provide a stringent comparison for LARS. We used 1mW to account for the write buffer leakage power.

To model a state-of-the-art embedded system microprocessor, we used configurations similar to the ARM Cortex A15 \cite{arm}. The microprocessor features a 2GHz clock frequency, separate 32KB STT-RAM L1 instruction and data caches, and an 8GB main memory. Based on our retention time analysis described in Section \ref{sec:miss}, we kept the L1 instruction cache's retention time at 100ms, while LARS was implemented in the data cache. We used a base retention time of 10ms for DRS, similar to prior work \cite{Jog12,Sun11}. Our LARS cache comprises of four STT-RAM units with 100$\mu$s, 1ms, 10ms, and 100ms retention times. We chose the retention times to be as low as possible without excessively increasing the cache miss rates with respect to the SRAM, while covering a range of application requirements. Note that the retention time is a design choice, and the cache can be designed with more STT-RAM units (and different retention times), depending on the design objectives and/or executing applications. However, we empirically determined that the chosen retention times suffice for the range of benchmarks used in our experiments.

To model different retention times, we used the MTJ cell size scaling method proposed in \cite{Chun13}. From the modeling results, we obtain essential parameters, such as the write pulse, write current, and MTJ resistance value $R_{AP}$. We then applied these parameters to the circuit modeling tool, NVSim \cite{NVSim}, in order to construct the STT-RAM cache for different retention times, as shown in Table \ref{tab:retention}. To fairly compare LARS with the SRAM cache, we also modeled the SRAM using NVSim. To represent a variety of workloads, we used twelve benchmarks from the SPEC 2006 benchmark suite compiled for the ARM instruction set architecture, using the reference input sets.  

\section{Results}\label{sec:results}

To illustrate LARS' effectiveness, we evaluated and analyzed the L1 data cache's energy and memory access latency achieved by LARS compared to the SRAM and DRS, representing prior work. In this section, we first summarize the results from the LARS sampling technique, and thereafter present results for the other LARS algorithms.

\begin{figure*}[ht]
\vspace{7pt}
\begin{subfigure}[t]{.5\textwidth}
  \centering
  \includegraphics[width=0.9\columnwidth]{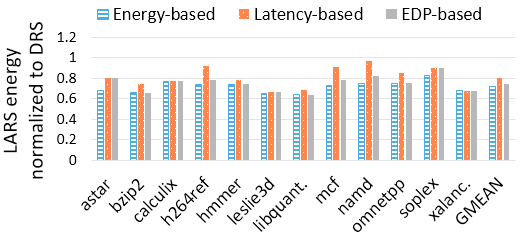}

  \caption{Energy normalized to DRS}
  \label{fig:exhaustive_energy}
\end{subfigure}%
~
\begin{subfigure}[t]{.5\textwidth}
  \centering
  \includegraphics[width=0.9\columnwidth]{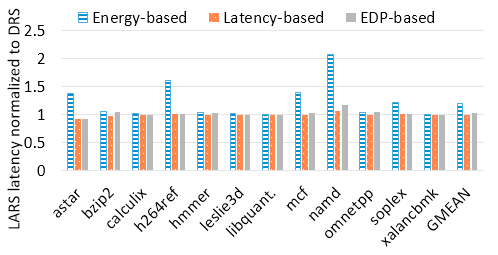}

  \caption{Latency normalized to DRS}
  \label{fig:exhaustive_latency}
\end{subfigure}

\caption{Energy and latency of LARS sampling technique normalized to DRS. Results shown for different objective functions: energy, latency, and EDP}
\label{fig:lars-optimal-approaches}
\end{figure*}

\subsection{Sampling Technique} \label{sec:samplingResult}
We evaluated the sampling technique based on different metrics---energy-, latency-, and EDP-based approaches---to determine which evaluation metric to use for the LARS tuning algorithm. Fig. \ref{fig:exhaustive_energy} illustrates the energy consumed by the different sampling approaches normalized to DRS. The energy-based approach achieved the highest energy reduction of 28.4\% on average across all the benchmarks, while the EDP- and latency-based approaches respectively reduced the energy by 25.31\% and 20.0\% on average. 
 
We also observed that LARS incurred some increases in cache misses as compared to DRS. This behavior resulted from the fact that unlike DRS where blocks are refreshed, LARS allows the blocks to expire when the current retention time as elapsed. However, for most applications, the increase in cache misses were not significant enough to cause a substantial increase in execution latency. Furthermore, smaller retention times enabled faster accesses, thus mitigating the overheads from the increase in cache misses. Some applications suffered substantial increases in misses (especially using the energy-based approach), leading to significant latency overheads compared to DRS.

Fig. \ref{fig:exhaustive_latency} depicts the latency achieved by LARS using the sampling technique for the energy-, latency, and EDP-based approaches normalized to DRS. Compared to DRS, on average, the energy-based approach traded off the latency for energy optimization, increasing the latency by 20.35\%, on average. The latency-based and EDP-based approaches, on the other hand, incurred marginal latency overheads of 0.04\% and 2.3\%, respectively. The substantial increase in latency for the energy-based approach was caused by a few benchmarks ($astar$, $h264ref$, and $namd$), which suffered substantial miss penalties in the process of reducing the energy. These trends led us to further analyze the results to understand the tradeoffs between the different approaches in order to minimize the overheads. Thus, we also explored the EDP impact of the different approaches.
 
The EDP-based approach reduced the EDP by 23.53\%, on average, while the latency- and energy-based approach reduced the EDP by 20.0\% and 13.82\%, respectively (graphs omitted for brevity). Even though the EDP-based approach increased the latency by 2.3\%, it substantially reduced both EDP and energy (both by over 20\%). While the EDP-based approach was not the best approach for either energy or latency, we found it to be the Pareto-optimal approach for a balanced tradeoff between energy and latency in our analysis. Therefore, we used the EDP-based approach as the objective function for the different LARS algorithms.

\begin{figure*}[t]
\begin{subfigure}[t]{.48\textwidth}
  \centering
  \includegraphics[width=0.7\columnwidth]{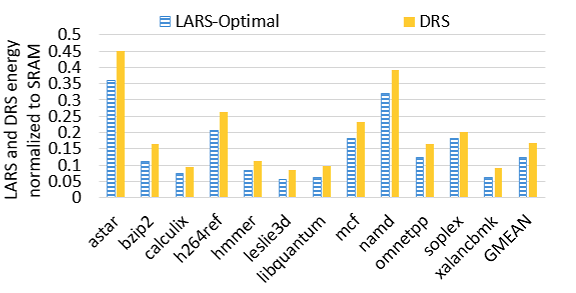}
  \caption{Data cache energy}
  \label{fig:energy_sram}
\end{subfigure}%
~
\begin{subfigure}[t]{.5\textwidth}
  \centering
  \includegraphics[width=0.7\columnwidth]{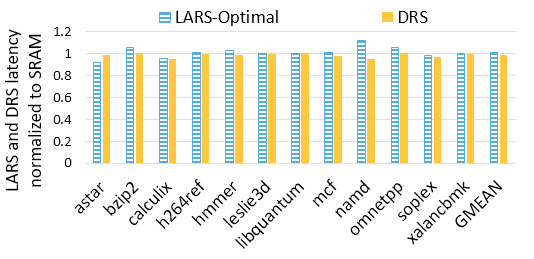}
  \caption{Data cache latency}
  \label{fig:latency_sram}
\end{subfigure}
\caption{LARS-Optimal and DRS data cache energy and latency normalized to SRAM}
\label{fig:cache_results_sram}
\end{figure*}

\subsection{LARS-Optimal Compared to the SRAM and DRS}\label{sec:normalSRAMResults}

Fig. \ref{fig:cache_results_sram} depicts the cache energy and latency achieved by both LARS-Optimal and DRS normalized to the SRAM. Fig. \ref{fig:energy_sram} shows that, on average across all the applications, LARS-Optimal reduced the energy by 87.56\% as compared to SRAM, with energy savings over 90\% for $calculix$, $hmmer$, $leslie3d$, $libquantum$, and $xalancbmk$. We note that this energy reduction was accounted for, in part, by the significantly reduced leakage power that STT-RAM offers as compared to SRAM (Table \ref{tab:retention}). Thus, both LARS-Optimal and DRS significantly reduced the energy compared to the SRAM. 

Our ultimate goal was to achieve energy improvements using LARS, without significantly degrading the latency. Hence, based on our analysis in Section \ref{sec:miss}, LARS uses the EDP as the objective function in order to minimize the latency expense of energy optimization. Fig. \ref{fig:latency_sram} shows that, on average, LARS-Optimal increased the latency by only 0.7\% as compared to the SRAM, with latency overheads as high as 11.5\% for $namd$. For a few benchmarks, such as $astar$, LARS \textit{reduced} the latency by up to 8.44\%. Even though LARS incurred some additional misses for some benchmarks, the additional misses were not enough to result in substantial latency overheads. The latency overheads were also mitigated, despite the increase in misses, by the fact that STT-RAM had faster read latency than SRAM (Table \ref{tab:retention}), as also observed in prior work \cite{Smullen11,DASCA}.

Compared to DRS, LARS-Optimal reduced the average energy by 25.31\%, with energy savings as high as 35.90\% for $libquantum$. LARS-Optimal was able to reduce the energy as compared to DRS by mitigating the energy overheads resulting from the dynamic refreshes featured in DRS. Furthermore, LARS-Optimal's improvement over DRS also resulted from LARS' ability to specialize the retention time to the executing application's requirements, unlike DRS, where a static retention time was used. As seen in Fig. \ref{fig:cache_results_sram}, LARS-Optimal outperformed DRS for all the benchmarks.

With respect to latency, LARS-Optimal \textit{increased} the latency by 2.3\%, on average, compared to DRS. For some benchmarks, however, LARS-Optimal reduced the latency by up to 7.0\% (for $astar$). We note that DRS outperformed LARS-Optimal---marginally---only because we modeled a "perfect" refresh scenario as described in section \ref{sec:setup}. The perfect scenario, which ignores the latency overheads of refreshes, may not always be the case in practice. Thus, these results are pessimistic for LARS-Optimal. We also reiterate that LARS-Optimal reduced the energy by 25.31\%, providing an appreciable energy improvement at the expense of some latency overhead.

In general, LARS' major advantage is that it adapts the retention time to different applications' runtime needs and uses a lower retention time when appropriate. In addition, LARS eliminated the need for dynamic refreshes, which was a source of overhead in DRS. We also observed that LARS performed best for applications that had short block lifetimes. That is, the applications cache miss rates remained low as the retention time reduced. $Leslie3d$ and $libquantum$ typify this behavior. As seen in Fig. \ref{fig:dcache_miss}, their cache misses remained low as the retention time reduced. Concomitantly, LARS' improvements over DRS were the highest two for these two applications---34.55\% and 35.90\% energy savings, respectively.

\begin{figure*}[t]
\begin{subfigure}[t]{.5\textwidth}
  \centering
  \includegraphics[width=0.7\columnwidth]{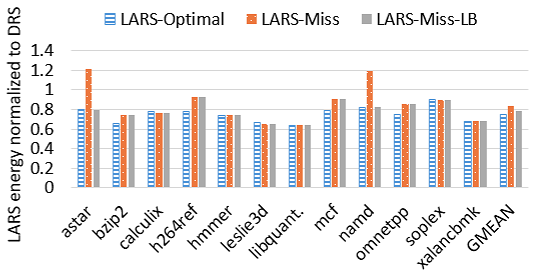}
  \vspace{-5pt}
  \caption{Data cache energy}
  \label{fig:energy_drs}
\end{subfigure}%
~
\begin{subfigure}[t]{.5\textwidth}
  \centering
  \includegraphics[width=0.7\columnwidth]{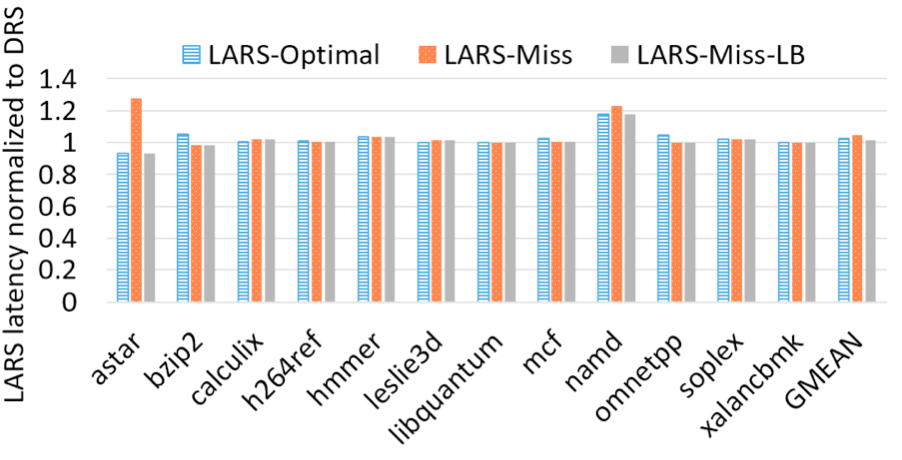}
  \vspace{-5pt}
  \caption{Data cache latency}
  \label{fig:latency_drs}
\end{subfigure}
\caption{Data cache energy and latency achieved by LARS (LARS-Optimal, LARS-Miss, and LARS-Miss-LB) normalized to DRS}
\label{fig:cache_results_drs}
\end{figure*}

\begin{figure}[ht]
    \centering
    \vspace{-5pt}
	\includegraphics[width=0.9\columnwidth]{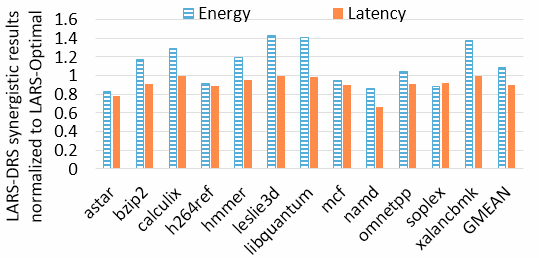}
	\caption{Energy and latency results of LARS-DRS synergistic scheme}
	\label{fig:synergistic}
\end{figure}

\subsection{LARS-Miss Compared to the DRS and LARS-Optimal}\label{sec:normalDRSResults}

Fig. \ref{fig:cache_results_drs} depicts the cache energy and latency achieved by LARS-Optimal, LARS-Miss, and LARS-Miss-LB normalized to DRS. On average across all the applications, LARS-Miss reduced the average energy by 16.68\%, with a latency overhead of 4.56\%, as compared to DRS. LARS-Miss reduced the energy by up to 35.9\% for $libquantum$. Compared to LARS-Optimal, LARS-Miss achieved similar or close energy results. However, two benchmarks---$astar$ and $namd$---were outliers with respect to LARS-Miss' performance. For these two benchmarks, we observed that their base number of misses were extremely small, and even though reducing the retention time would have increased the number of misses, this increase was not significant enough to translate to increased dynamic energy as described in our prior analysis (Section \ref{sec:algorithm}). However, LARS-Miss selected the 100ms retention time for these two benchmarks, resulting in much higher energy consumption and longer latency.

We observed that LARS-Miss-LB's performance with respect to energy was closer to LARS-Optimal than LARS-Miss. Compared to DRS, LARS-Miss-LB reduced the average energy by 21.96\%, whereas LARS-Optimal and LARS-Miss reduced the average energy by 25.31\% and 16.68\%, respectively. As shown in Fig. \ref{fig:energy_drs}, LARS-Miss-LB outperformed DRS for all the benchmarks. For instance, even though DRS outperformed LARS-Miss for $astar$ and $namd$, LARS-Miss-LB reduced the energy for these two benchmarks by 41.2\% and 37.0\%, respectively. Furthermore, LARS-Miss-LB also reduced the latency overhead to 1.4\%, from the 4.56\% and 2.3\% latency overhead of LARS-Miss and LARS-Optimal, respectively. The most important feature of both LARS-Miss and LARS-Miss-LB is that they eliminated the need for the more complex energy calculation circuits present in LARS-Optimal. By eliminating the need for a substantial part of the tuner datapath, these algorithms traded off search accuracy in favor of reduced hardware overhead and static energy.

\subsection{Exploring a Synergy Between LARS and DRS}
\label{sec:overhead}
We also explored the benefits in combining LARS and DRS in order to achieve additional energy savings. We implemented a synergistic scheme that featured the best retention time (equivalent to LARS-Optimal) and a refresh mechanism to prevent premature data evictions (equivalent to DRS). Fig. \ref{fig:synergistic} depicts the energy and latency achieved by this synergistic scheme normalized to LARS-Optimal. The synergy of LARS and DRS reduced the average latency by 9.82\% as compared to LARS-Optimal, with latency reduction of up to 33.89\% for $namd$. However, the synergistic scheme increased energy by 9.06\% on average, and substantially increased the energy in several benchmarks. For example, the average energy compared to LARS-Optimal was increased by 42.96\%, 40.47\%, and 37.33\% for $leslie3d$, $libquantum$, and $xalancbmk$, respectively. We attribute these results to the fact that the synergistic scheme required a buffer to enable the refresh operations. Since the buffer contributed leakage and dynamic power, the benchmarks with longer latency suffered substantial energy degradation. As such, $leslie3d$, $libquantum$, and $xalancbmk$, which were the longest benchmarks in our experiments, exhibited the highest energy increase. We observed only marginal EDP improvements from this synergistic scheme as compared to LARS-Optimal. On average, the synergistic scheme only improved the EDP by 1.65\%.

\section{Conclusions and Future Research}
In this paper, we explored the applicability of dynamic retention times in both instruction and data STT-RAM L1 caches. Our analysis revealed that while static retention times suffice for the instruction cache, much energy benefits can be derived from adapting the data cache's retention times to applications' variable runtime requirements, based on the applications' characteristics. 
To this end, we proposed \textit{LARS: Logically Adaptable Retention Time STT-RAM} cache, which logically adapts the STT-RAM's retention time to different applications' runtime requirements, in order to reduce the write energy, with minimal overheads. LARS comprises of multiple STT-RAM units with different retention times; only one unit is used at a time, depending on an application's needs. Based on our analysis of applications' characteristics with respect to the retention time and the LARS cache architecture, we proposed tuning algorithms to determine the best retention time at runtime. Experiments show that LARS can reduce the average energy by up to 25.31\%, as compared to prior related work, without incurring significant latency or area overheads. 

For future work, we intend to explore finer grained optimizations by exploring LARS' impact for applications' dynamic runtime phases. We will also explore the synergy of LARS with the adaptability of other cache parameters (cache size, line size, associativity, replacement policy), in order to fully satisfy executing applications' resource requirements.   
	
	\bibliographystyle{IEEEtran}
	\bibliography{refs}

\begin{thebibliography}{10}
\providecommand{\url}[1]{#1}
\csname url@samestyle\endcsname
\providecommand{\newblock}{\relax}
\providecommand{\bibinfo}[2]{#2}
\providecommand{\BIBentrySTDinterwordspacing}{\spaceskip=0pt\relax}
\providecommand{\BIBentryALTinterwordstretchfactor}{4}
\providecommand{\BIBentryALTinterwordspacing}{\spaceskip=\fontdimen2\font plus
\BIBentryALTinterwordstretchfactor\fontdimen3\font minus
  \fontdimen4\font\relax}
\providecommand{\BIBforeignlanguage}[2]{{%
\expandafter\ifx\csname l@#1\endcsname\relax
\typeout{** WARNING: IEEEtran.bst: No hyphenation pattern has been}%
\typeout{** loaded for the language `#1'. Using the pattern for}%
\typeout{** the default language instead.}%
\else
\language=\csname l@#1\endcsname
\fi
#2}}
\providecommand{\BIBdecl}{\relax}
\BIBdecl

\bibitem{mittal14}
S.~Mittal, ``A survey of architectural techniques for improving cache power
  efficiency,'' \emph{Sustainable Computing: Informatics and Systems}, vol.~4,
  no.~1, pp. 33--43, 2014.

\bibitem{Jog12}
A.~Jog, A.~K. Mishra, C.~Xu, Y.~Xie, V.~Narayanan, R.~Iyer, and C.~R. Das,
  ``Cache revive: Architecting volatile stt-ram caches for enhanced performance
  in cmps,'' in \emph{DAC Design Automation Conference 2012}, June 2012, pp.
  243--252.

\bibitem{Dong08}
\BIBentryALTinterwordspacing
X.~Dong, X.~Wu, G.~Sun, Y.~Xie, H.~Li, and Y.~Chen, ``Circuit and
  microarchitecture evaluation of 3d stacking magnetic ram (mram) as a
  universal memory replacement,'' in \emph{Proceedings of the 45th Annual
  Design Automation Conference}, ser. DAC '08.\hskip 1em plus 0.5em minus
  0.4em\relax New York, NY, USA: ACM, 2008, pp. 554--559. [Online]. Available:
  \url{http://doi.acm.org/10.1145/1391469.1391610}
\BIBentrySTDinterwordspacing

\bibitem{Sun11}
Z.~Sun, X.~Bi, H.~Li, W.~F. Wong, Z.~L. Ong, X.~Zhu, and W.~Wu, ``Multi
  retention level stt-ram cache designs with a dynamic refresh scheme,'' in
  \emph{2011 44th Annual IEEE/ACM International Symposium on Microarchitecture
  (MICRO)}, Dec 2011, pp. 329--338.

\bibitem{apalkov13}
D.~Apalkov, A.~Khvalkovskiy, S.~Watts, V.~Nikitin, X.~Tang, D.~Lottis, K.~Moon,
  X.~Luo, E.~Chen, A.~Ong \emph{et~al.}, ``Spin-transfer torque magnetic random
  access memory (stt-mram),'' \emph{ACM Journal on Emerging Technologies in
  Computing Systems (JETC)}, vol.~9, no.~2, p.~13, 2013.

\bibitem{chung16}
S.-W. Chung, T.~Kishi, J.~Park, M.~Yoshikawa, K.~Park, T.~Nagase, K.~Sunouchi,
  H.~Kanaya, G.~Kim, K.~Noma \emph{et~al.}, ``4gbit density stt-mram using
  perpendicular mtj realized with compact cell structure,'' in \emph{Electron
  Devices Meeting (IEDM), 2016 IEEE International}.\hskip 1em plus 0.5em minus
  0.4em\relax IEEE, 2016, pp. 27--1.

\bibitem{park12}
S.~P. Park, S.~Gupta, N.~Mojumder, A.~Raghunathan, and K.~Roy, ``Future cache
  design using stt mrams for improved energy efficiency: devices, circuits and
  architecture,'' in \emph{Proceedings of the 49th Annual Design Automation
  Conference}.\hskip 1em plus 0.5em minus 0.4em\relax ACM, 2012, pp. 492--497.

\bibitem{Diao07}
\BIBentryALTinterwordspacing
Z.~Diao, Z.~Li, S.~Wang, Y.~Ding, A.~Panchula, E.~Chen, L.-C. Wang, and
  Y.~Huai, ``Spin-transfer torque switching in magnetic tunnel junctions and
  spin-transfer torque random access memory,'' \emph{Journal of Physics:
  Condensed Matter}, vol.~19, no.~16, p. 165209, 2007. [Online]. Available:
  \url{http://stacks.iop.org/0953-8984/19/i=16/a=165209}
\BIBentrySTDinterwordspacing

\bibitem{CongXu11}
C.~Xu, D.~Niu, X.~Zhu, S.~H. Kang, M.~Nowak, and Y.~Xie, ``Device-architecture
  co-optimization of stt-ram based memory for low power embedded systems,'' in
  \emph{2011 IEEE/ACM International Conference on Computer-Aided Design
  (ICCAD)}, Nov 2011, pp. 463--470.

\bibitem{Smullen11}
C.~W. Smullen, V.~Mohan, A.~Nigam, S.~Gurumurthi, and M.~R. Stan, ``Relaxing
  non-volatility for fast and energy-efficient stt-ram caches,'' in \emph{2011
  IEEE 17th International Symposium on High Performance Computer Architecture},
  Feb 2011, pp. 50--61.

\bibitem{Li13}
Q.~Li, J.~Li, L.~Shi, C.~J. Xue, Y.~Chen, and Y.~He, ``Compiler-assisted
  refresh minimization for volatile stt-ram cache,'' in \emph{2013 18th Asia
  and South Pacific Design Automation Conference (ASP-DAC)}, Jan 2013, pp.
  273--278.

\bibitem{Rodriguez14}
\BIBentryALTinterwordspacing
G.~Rodr\'{\i}guez, J.~Touri\~{n}o, and M.~T. Kandemir, ``Volatile stt-ram
  scratchpad design and data allocation for low energy,'' \emph{ACM Trans.
  Archit. Code Optim.}, vol.~11, no.~4, pp. 38:1--38:26, Dec. 2014. [Online].
  Available: \url{http://doi.acm.org/10.1145/2669556}
\BIBentrySTDinterwordspacing

\bibitem{Qiu16}
K.~Qiu, J.~Luo, Z.~Gong, W.~Zhang, J.~Wang, Y.~Xu, T.~Li, and C.~J. Xue,
  ``Refresh-aware loop scheduling for high performance low power volatile
  stt-ram,'' in \emph{2016 IEEE 34th International Conference on Computer
  Design (ICCD)}, Oct 2016, pp. 209--216.

\bibitem{zhang03}
C.~Zhang, F.~Vahid, and W.~Najjar, ``A highly configurable cache architecture
  for embedded systems,'' in \emph{Computer Architecture, 2003. Proceedings.
  30th Annual International Symposium on}.\hskip 1em plus 0.5em minus
  0.4em\relax IEEE, 2003, pp. 136--146.

\bibitem{Chun13}
K.~C. Chun, H.~Zhao, J.~D. Harms, T.~H. Kim, J.~P. Wang, and C.~H. Kim, ``A
  scaling roadmap and performance evaluation of in-plane and perpendicular mtj
  based stt-mrams for high-density cache memory,'' \emph{IEEE Journal of
  Solid-State Circuits}, vol.~48, no.~2, pp. 598--610, Feb 2013.

\bibitem{Bouziane18}
\BIBentryALTinterwordspacing
R.~Bouziane, E.~Rohou, and A.~Gamati{\'e}, ``Compile-time silent-store
  elimination for energy efficiency: An analytic evaluation for non-volatile
  cache memory,'' in \emph{Proceedings of the Rapido'18 Workshop on Rapid
  Simulation and Performance Evaluation: Methods and Tools}, ser. RAPIDO
  '18.\hskip 1em plus 0.5em minus 0.4em\relax New York, NY, USA: ACM, 2018, pp.
  5:1--5:8. [Online]. Available:
  \url{http://doi.acm.org/10.1145/3180665.3180666}
\BIBentrySTDinterwordspacing

\bibitem{Ranjan17}
A.~Ranjan, S.~Venkataramani, Z.~Pajouhi, R.~Venkatesan, K.~Roy, and
  A.~Raghunathan, ``Staxcache: An approximate, energy efficient stt-mram
  cache,'' in \emph{Design, Automation Test in Europe Conference Exhibition
  (DATE), 2017}, March 2017, pp. 356--361.

\bibitem{Reed17}
\BIBentryALTinterwordspacing
E.~Reed, A.~R. Alameldeen, H.~Naeimi, and P.~Stolt, ``Probabilistic replacement
  strategies for improving the lifetimes of nvm-based caches,'' in
  \emph{Proceedings of the International Symposium on Memory Systems}, ser.
  MEMSYS '17.\hskip 1em plus 0.5em minus 0.4em\relax New York, NY, USA: ACM,
  2017, pp. 166--176. [Online]. Available:
  \url{http://doi.acm.org/10.1145/3132402.3132433}
\BIBentrySTDinterwordspacing

\bibitem{RR17}
\BIBentryALTinterwordspacing
R.~Rodríguez-Rodríguez, J.~Díaz, F.~Castro, P.~Ibáñez, D.~Chaver,
  V.~Viñals, J.~C. Saez, M.~Prieto-Matias, L.~Piñuel, T.~Monreal, and J.~M.
  Llabería, ``Reuse detector: Improving the management of stt-ram sllcs,''
  \emph{The Computer Journal}, pp. 1--25, 2017. [Online]. Available: \url{+
  http://dx.doi.org/10.1093/comjnl/bxx099}
\BIBentrySTDinterwordspacing

\bibitem{DASCA}
J.~Ahn, S.~Yoo, and K.~Choi, ``Dasca: Dead write prediction assisted stt-ram
  cache architecture,'' in \emph{2014 IEEE 20th International Symposium on High
  Performance Computer Architecture (HPCA)}, Feb 2014, pp. 25--36.

\bibitem{Benzene}
\BIBentryALTinterwordspacing
N.~Kim, J.~Ahn, K.~Choi, D.~Sanchez, D.~Yoo, and S.~Ryu, ``Benzene: An
  energy-efficient distributed hybrid cache architecture for manycore
  systems,'' \emph{ACM Trans. Archit. Code Optim.}, vol.~15, no.~1, pp.
  10:1--10:23, Mar. 2018. [Online]. Available:
  \url{http://doi.acm.org/10.1145/3177963}
\BIBentrySTDinterwordspacing

\bibitem{Korgaonkar18}
K.~Korgaonkar, I.~Bhati, H.~Liu, J.~Gaur, S.~Manipatruni, S.~Subramoney,
  T.~Karnik, S.~Swanson, I.~Young, and H.~Wang, ``Density tradeoffs of
  non-volatile memory as a replacement for sram based last level cache,'' in
  \emph{2018 ACM/IEEE 45th Annual International Symposium on Computer
  Architecture (ISCA)}, June 2018, pp. 315--327.

\bibitem{Zeng17}
Q.~Zeng and J.~K. Peir, ``Content-aware non-volatile cache replacement,'' in
  \emph{2017 IEEE International Parallel and Distributed Processing Symposium
  (IPDPS)}, May 2017, pp. 92--101.

\bibitem{spec2006}
\BIBentryALTinterwordspacing
J.~L. Henning, ``Spec cpu2006 benchmark descriptions,'' \emph{SIGARCH Comput.
  Archit. News}, vol.~34, no.~4, pp. 1--17, Sep. 2006. [Online]. Available:
  \url{http://doi.acm.org/10.1145/1186736.1186737}
\BIBentrySTDinterwordspacing

\bibitem{Kang15}
W.~Kang, L.~Zhang, W.~Zhao, J.~Klein, Y.~Zhang, D.~Ravelosona, and C.~Chappert,
  ``Yield and reliability improvement techniques for emerging nonvolatile
  stt-mram,'' \emph{IEEE Journal on Emerging and Selected Topics in Circuits
  and Systems}, vol.~5, no.~1, pp. 28--39, March 2015.

\bibitem{Kang17}
W.~Kang, L.~Chang, Z.~Wang, W.~Lv, G.~Sun, and W.~Zhao, ``Pseudo-differential
  sensing framework for stt-mram: A cross-layer perspective,'' \emph{IEEE
  Transactions on Computers}, vol.~66, no.~3, pp. 531--544, March 2017.

\bibitem{ando14}
K.~Ando, S.~Fujita, J.~Ito, S.~Yuasa, Y.~Suzuki, Y.~Nakatani, T.~Miyazaki, and
  H.~Yoda, ``Spin-transfer torque magnetoresistive random-access memory
  technologies for normally off computing,'' \emph{Journal of Applied Physics},
  vol. 115, no.~17, p. 172607, 2014.

\bibitem{Tosi15}
\BIBentryALTinterwordspacing
T.~Adegbija and A.~Gordon-Ross, ``Phase-based cache locking for embedded
  systems,'' in \emph{Proceedings of the 25th Edition on Great Lakes Symposium
  on VLSI}, ser. GLSVLSI '15.\hskip 1em plus 0.5em minus 0.4em\relax New York,
  NY, USA: ACM, 2015, pp. 115--120. [Online]. Available:
  \url{http://doi.acm.org/10.1145/2742060.2742076}
\BIBentrySTDinterwordspacing

\bibitem{gordon-ross07}
A.~Gordon-Ross and F.~Vahid, ``A self-tuning configurable cache,'' in
  \emph{Proceedings of the 44th annual Design Automation Conference}.\hskip 1em
  plus 0.5em minus 0.4em\relax ACM, 2007, pp. 234--237.

\bibitem{synopsys}
D.~Compiler, ``Synopsys inc,'' 2000.

\bibitem{CortexA15area}
\BIBentryALTinterwordspacing
F.~A. Endo, D.~Courouss{\'e}, and H.-P. Charles, ``Micro-architectural
  simulation of embedded core heterogeneity with gem5 and mcpat,'' in
  \emph{Proceedings of the 2015 Workshop on Rapid Simulation and Performance
  Evaluation: Methods and Tools}, ser. RAPIDO '15.\hskip 1em plus 0.5em minus
  0.4em\relax New York, NY, USA: ACM, 2015, pp. 7:1--7:6. [Online]. Available:
  \url{http://doi.acm.org/10.1145/2693433.2693440}
\BIBentrySTDinterwordspacing

\bibitem{gem5}
\BIBentryALTinterwordspacing
N.~Binkert, B.~Beckmann, G.~Black, S.~K. Reinhardt, A.~Saidi, A.~Basu,
  J.~Hestness, D.~R. Hower, T.~Krishna, S.~Sardashti, R.~Sen, K.~Sewell,
  M.~Shoaib, N.~Vaish, M.~D. Hill, and D.~A. Wood, ``The gem5 simulator,''
  \emph{SIGARCH Comput. Archit. News}, vol.~39, no.~2, pp. 1--7, Aug. 2011.
  [Online]. Available: \url{http://doi.acm.org/10.1145/2024716.2024718}
\BIBentrySTDinterwordspacing

\bibitem{arm}
\BIBentryALTinterwordspacing
``{Cortex-A15 Processor}.'' [Online]. Available:
  \url{https://www.arm.com/products/processors/cortex-a/cortex-a15.php}
\BIBentrySTDinterwordspacing

\bibitem{NVSim}
\BIBentryALTinterwordspacing
X.~Dong, C.~Xu, Y.~Xie, and N.~P. Jouppi, ``Nvsim: A circuit-level performance,
  energy, and area model for emerging nonvolatile memory,'' \emph{Trans.
  Comp.-Aided Des. Integ. Cir. Sys.}, vol.~31, no.~7, pp. 994--1007, Jul. 2012.
  [Online]. Available: \url{http://dx.doi.org/10.1109/TCAD.2012.2185930}
\BIBentrySTDinterwordspacing

\end{thebibliography}

\vspace{120pt}
\begin{IEEEbiography}[{\includegraphics[width=1in,height=1.25in,clip,keepaspectratio]{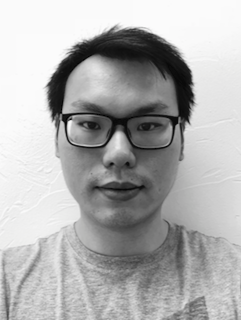}}]{Kyle Kuan} (M'16) is a Ph.D. student in the Department of Electrical and Computer Engineering at the University of Arizona. He received his M.S. in Electrical Engineering from National Taiwan University in 2008 and B.S. in Mechanical Engineering from National Chiao Tung University in 2006. His research interests include cache design for energy efficient systems, non-volatile memories, and right-provisioned micro architectures for IoT devices.
\end{IEEEbiography}
\vspace{-400pt}
\begin{IEEEbiography}[{\includegraphics[width=1in,height=1.25in,clip,keepaspectratio]{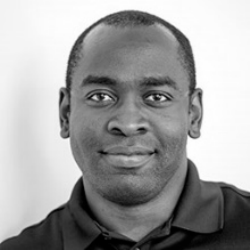}}]{Tosiron Adegbija} (M'11) received his M.S and Ph.D in Electrical and Computer Engineering from the University of Florida in 2011 and 2015, respectively and his B.Eng in Electrical Engineering from the University of Ilorin, Nigeria in 2005. 

He is currently an Assistant Professor of Electrical and Computer Engineering at the University of Arizona, USA. His research interests are in computer architecture, with emphasis on adaptable computing, low-power embedded systems design and optimization methodologies, and microprocessor optimizations for the Internet of Things (IoT). 

Dr. Adegbija was a recipient of the Best Paper Award at the Ph.D forum of IEEE Computer Society Annual Symposium on VLSI (ISVLSI) in 2014.
\end{IEEEbiography}

\balance
\end{document}